\documentclass{elsart}
\usepackage{graphicx}
\usepackage{bm}
\usepackage{epsfig}
\usepackage{amssymb}
\usepackage{amsmath}
\newcommand{\ve}{\varepsilon}

\begin{document}
\begin{frontmatter}
\title{Chaos in a generalized Lorenz system}
\author{E.D.~Belokolos}%
\ead{bel@imag.kiev.ua}%
\author{V.O.~Kharchenko}%
\ead{vasiliy@imag.kiev.ua}%
\address{Institute of Magnetism, Nat. Acad. Sci. of Ukraine \break
36-B, Vernadsky St., 03680 Kyiv, Ukraine}
\author{D.O.~Kharchenko}%
\ead{dikh@ipfcentr.sumy.ua}%
\address{Applied Physics Institute, Nat. Acad. Sci. of Ukraine \break
56, Petropavlovskaya St., 40030 Sumy, Ukraine}

\date{}

\begin{abstract}
A three-component dynamic system with influence of pumping and
nonlinear dissipation describing a quantum cavity electrodynamic
device is studied. Different dynamical regimes are investigated in
terms of divergent trajectories approaches and fractal statistics.
It has been shown, that in such a system stable and unstable
dissipative structures type of limit cycles can be formed with
variation of pumping and nonlinear dissipation rate. Transitions
to chaotic regime and the corresponding chaotic attractor are
studied in details.
\end{abstract}


\end{frontmatter}

\section{Introduction}

The most intriguing phenomenon in nonlinear dynamical systems theory is a
transition from regular dynamics to irregular one. Despite a transition to
irregular dynamics can be driven by stochastic sources introduction into
corresponding evolution equations, a special interest from the theoretical
viewpoint is an observation of such regime in deterministic systems. A famous
work by E.Lorenz \cite{Lorenz} shows a possibility of regular dynamical systems
to exhibit chaotic regime. After, a lot of works concerning this problem
allowed to develop mathematical tools to study chaotic regime in different kind
of physical, chemical, biological and other systems, see
Refs.\cite{Zaslavskiy,shuster,yellow_book,bokaletti}, for example.

Mathematically studying of Lorenz system was performed in \cite{math1,math11},
where authors introduced the geometric Lorenz flows. Invariance principle for
geometric Lorenz attractors were obtained in \cite{math2}. Description of
dynamics of the Lorenz system having an invariant algebraic surface presented
in \cite{math3}.

In physics, thermodynamics in particular, chaos is applied in the study of
turbulence leading to the understanding of self-organizing systems and system
states (equilibrium, near equilibrium, the edge of chaos, and chaos). The
occurrence of a chaotic regime is related to the interplay between the
instability and nonlinearity. The instability is responsible for the
exponential divergence of two nearby trajectories, the nonlinearity bounds
trajectories within the finite volume of the phase space of the system. The
combination these two mechanisms gives rise to a high sensitivity of the system
to the initial conditions.

It is well known, that optoelectronical devices are the simplest systems which
manifest chaotic regime under deterministic conditions. In practice, a
controlling of dynamical regimes in optoelectronical devices is very actual
task, because different types of radiation can be used for different areas of
human activity \cite{Haken80,Hanin}. The problem to control chaotic regimes in
laser dynamics attracts much more attention in last decades. Nowadays new
mechanisms which lead to chaotic regime generation were found
\cite{ieejournal,6fromiee,7fromiee}. Moreover, a rapidly developing approach to
study the quantum cavity electrodynamic models shows a possibility of chaotic
radiation and the chaos control \cite{arhiv,14z}. From a laser physics it
follows that self-organization processes with dissipative structures formation
or generation of chaotic signal can be stimulated by introducing additional
nonlinear medium into the Fabry-Perot cavity
\cite{kachmarek,Hanin,Haken,Lugiato1974,GETF71,Optics2002}. Experimental
studies of the chaotic regime in such systems allows to generalize well known
theoretical models, identifying additional nonlinearity sources, in order to
provide a more detailed theoretical analysis and to predict different types of
laser dynamics \cite{sdarticle4,numeric2002,exp2003,zigzag}.

From the theoretical viewpoint, a problem to find ability of the system to
manifest chaotic regimes and the control of chaos is of a special interest in a
nonlinear dynamical system theory. Moreover, due to chaotic dynamics is
accompanied by a different type of attractors appearing, another interesting
problem is to investigate properties of such attractors. On physical grounds,
the related actual problem is to investigate a chaotic regime in a generalized
model of a dynamical system, whereas an initial model does not exhibits it
itself.

In this paper we solve above problems for a three-parametric dynamical system
realized in optoelectronical devices. We are aimed to study a possibility of
the well known two-level laser system type of Lorenz-Haken model
\cite{Haken80,Haken}, which can not show chaotic regime in actual ranges of
system parameters, but manifests it if an additional nonlinear medium is
introduced into the cavity. In our investigation we use suppositions of
absorptive optical bistability and consider three-parametric system in order to
show that a transition to chaotic regime can be controlled by the pumping and a
bistability parameter proportional to the atomic density. The last one is
related to nonlinear dependence of a relaxation time of one of dynamical
variables versus its value. It will be shown that varying in a system
parameters one can arrive at stable and unstable dissipative structures, period
doubling bifurcation and a chaotic regime. A corresponding chaotic attractor is
studied by different approaches.

The paper is organized in the following manner. In Section 2 we present a model
of our system where we introduce theoretical constructions to model an
influence of a nonlinear dissipation. Section 3 is devoted to development of
the analytical approach to study possible regimes. In Section 4 we apply the
derived formalism to investigate a reconstruction of the dynamics of the
system. Section 5 is devoted to considering the main properties of the chaotic
behaviour of a system. Main results and perspectives are collected in the
Conclusions.

\section{Model}

Let us consider a three parametric dynamical model, which is widely used to
describe self-organization processes:
\begin{equation}
\left\{
\begin{split}
 &\dot{x}=-x+y,\\
 &\sigma\dot{y}=-y+xz,\\
 &\ve\dot{z}=(r-z)-xy,
\end{split}
\right. \label{lor2eq}
\end{equation}
where dot means derivative respect to the time $t$. This model can be simply
derived from the well known chaotic Lorenz system \cite{Lorenz} using following
relations: $t'=\sigma t$, $x=X/\sqrt{b}$, $y=Y/\sqrt{b}$, $z=r-Z$,
$\ve=\sigma/b$. Here $X$, $Y$, $Z$ are dynamical variables of the Lorenz
system; $\sigma$, $b$ and $r$ are related constants; we drop the prime in the
time $t$ for convenience. The dynamical system (\ref{lor2eq}) can be obtained
in the lasers theory where the Maxwell-type equations for electro-magnetic
field and a density matrix evolution equation are exploited. It leads to the
system of the Maxwell-Bloch type that is reduced to the Lorenz-Haken model
(\ref{lor2eq}) for two level laser system (see, for example,
Refs.\cite{Haken,Hanin,Gardiner2000}). In such a case $x$, $y$, $z$ are
addressed to a strength of the electric field, polarization and an inversion
population of energy levels. Models type of Eq.(\ref{lor2eq}) are extensively
used to describe synergetic transitions in complex systems
\cite{olemskoy_book}, where above variables $x$, $y$, $z$ acquire a physical
meaning of an order parameter, a conjugated field and a control parameter; the
quantity $r$ relates to the pump intensity and measures influence of the
environment.

A naive consideration of dynamical regimes in the model Eq.(\ref{lor2eq}) at
$r>0$ and $\sigma\simeq\ve\simeq 1$ shows that despite the system has
nonlinearities in both second and third equations its behaviour is well
defined. In this paper we consider a more general case reduced to the
absorptive optical bistability models \cite{Lugiato1974,Haken}. From the
physical viewpoint absorptive optical bistability is related to possibility of
the additional medium in the Fabry-Perot cavity (phthalocyanine fluid
\cite{Optics2002}, gases $SF_6$, $BaCl_3$ and $CO_2$ \cite{GETF71,Haken}) to
transit a radiation with high intensities only with absorption of week signals.
Despite models of such a type were considered in the one-parametric case and
only a stationary picture of dynamical system behaviour is analyzed (see
Ref.\cite{book_sumy} and citations therein), we will focus onto describing
possible dynamical regimes in the three-parametric model. Mathematically, our
model reads
\begin{equation}
\left\{
\begin{split}
 &\dot{x}=-x+y+f_\kappa, \quad f_\kappa=-\frac{\kappa x}{1+x^2},\\
 &\sigma\dot{y}=-y+xz,\\
 &\ve\dot{z}=(r-z)-xy.
\end{split}
\right. \label{lor3eq}
\end{equation}
Here the additional force $f_\kappa$ is introduced to take into account
influence of an additional medium; $\kappa$ is the so-called bistability
parameter proportional to the atomic density. Formally, it means a dispersion
of a relaxation time for the variable $x$. Indeed, if we rewrite the first
equation in the form $\dot{x}=-x/\tau(x)+y$, then the dispersive relaxation
time is $\tau(x)=1-\kappa/(1+\kappa+x^2)$.

Our main goal is to study possible scenarios of a dynamical regimes
reconstruction in such three-parametric system under influence of the pumping
$r$ and the nonlinear dissipation controlled by $\kappa$. We will investigate
conditions of a chaotic regime appearance. The corresponding irregular dynamics
well be studied in details. It will be shown that a chaotic regime can be
controlled by pumping and properties of an additional medium.

\section{Stability analysis}

Let us consider conditions where all variables are commensurable, setting
$\sigma,\epsilon\sim 1$. A behaviour of the system we present in the phase
space $(x,y,z)$. At first, let us study steady states. Setting $\dot x=0$,
$\dot y=0$ and $\dot z=0$, one can find fixed points described by the values
$x_0^{(k)}$, $y_0^{(k)}$ and $z_0^{(k)}$, where $k=1,2,3$. The first one has
coordinates
\begin{equation}
\left(x_0^{(1)},y_0^{(1)},z_0^{(1)}\right)=(0,0,r)\label{std_sts_1}
\end{equation}
and exists always; two symmetrical points, denoted with superscripts 2 and 3
respectively, have coordinates
\begin{equation}
\begin{split}
 &\left(x_0^{(2,3)},y_0^{(2,3)},z_0^{(2,3)}\right)=\\
 &\left(\pm\sqrt{r-\kappa-1},\frac{\pm r\sqrt{r-\kappa-1}}{r-\kappa},\frac{r}{r-\kappa}\right)
\end{split}
\label{std_sts_2}
\end{equation}
and realized under the condition
\begin{equation}
r\geq\kappa+1. \label{cond_single}
\end{equation}
It means, that $r_c=\kappa+1$ is a critical point for the bifurcation of
doubling.

According to the stationary states behaviour under influence of pumping and
nonlinear dissipation, let us perform a local linear stability analysis
\cite{Poincare,Andronov}. Within the framework of the standard Lyapunov
exponents approach time depended solutions of the system (\ref{lor3eq}) are
assumed in the form $\vec{u}\propto e^{\Lambda t},\quad
\Lambda=\lambda+i\omega$, where $\vec{u}\equiv{(x,y,z)}$, $\lambda$ controls
the stability of the stationary points, $\omega$ determines a frequency, as
usual. Magnitudes for real and imaginary parts of $\Lambda$ are calculated
according to the Jacobi matrix elements
$M_{ij}\equiv\left({\partial{f^{(i)}}}/{\partial{u_j}}\right)_{u_j=u_{j0}}$,
$i,j=x,y,z$; $f^{(i)}$ represents the right hand side of the corresponding
dynamical equations in Eq.(\ref{lor3eq}); the subscript 0 relates to the
stationary value. As a result, the Jacobi matrix takes the form
\begin{equation}
 M=-
 \left(
 \begin{split}
 1+\kappa\frac{1-x_0^2}{(1+x_0^2)^2}+\Lambda &\ & -1 &\ & 0\\
 -z_0 &\ & 1+\Lambda &\ & -x_0\\
 y_0 &\ & x_0 &\ & 1+\Lambda
 \end{split}
 \right).
\end{equation}

Let us define the stability of the fixed point (\ref{std_sts_1}). A solution of
the eigenvalue problem yields that the three eigenvalues are real and negative:
\begin{equation}
\begin{split}
 & \Lambda_1=\lambda_1=-1 <0,\\
 & \Lambda_2=\lambda_2=-\frac{1}{2}\kappa-1-\frac{1}{2}\sqrt{\kappa^2+4r} <0,\\
 & \Lambda_3=\lambda_3=-\frac{1}{2}\kappa-1+\frac{1}{2}\sqrt{\kappa^2+4r} <0.
\end{split}
\label{lambda_i}
\end{equation}
It means that if both $\kappa$ and $r$ satisfy the condition
(\ref{cond_single}), then the phase space is characterized by only node
(\ref{std_sts_1}).

On the other hand, at $r>r_c$, all three fixed points are realized in the phase
space. As it follows from our analysis the fixed point ({\ref{std_sts_1})
changes its stability and becomes a saddle, due to $\lambda_3>0$.

The stability of other two points with coordinates given by
Eq.(\ref{std_sts_2}) can be set from solutions of the cubic equation
\begin{equation}
\begin{split}
 \Lambda^3+&\left[3+\kappa\frac{2+\kappa-r}{(r-\kappa)^2}\right]\Lambda^2\\
 +&\left[2+\kappa-r-\frac{r}{r-\kappa}+2\kappa\frac{2+\kappa-r}{(r-\kappa)^2}\right]\Lambda\\
 +&\left[2+\kappa+\kappa\left(\frac{2+\kappa-r}{r-\kappa}\right)^2\right]=0.
 \label{Lambda_3}
\end{split}
\end{equation}

Analytically, we can define only a border of the domain of system parameters
where periodic solutions are realized. To consider a possibility of limit
cycles formation in the phase space let us assume $\Lambda=\pm{\rm i}\omega$
and insert it into Eq(\ref{Lambda_3}). It yields the equation for such a border
in the form
\begin{equation}
\begin{split}
 &2\kappa\frac{2+\kappa-r}{(r-\kappa)^2}\left[\kappa-r-6+\frac{r}{2}\frac{1}{r-\kappa}\right]\\
 -&2\kappa^2\left(\frac{2+\kappa-r}{(r-\kappa)^2}\right)^2\\
 +&4(\kappa-1)+3r\left(\frac{1}{r-\kappa}-1\right)=0.
\label{eq19}
\end{split}
\end{equation}
As it will be seen below inside this domain in the plane $(r,\kappa)$ only
stable periodic solutions are realized.

\section{Analysis of dynamical regimes}

To describe the dynamical regimes of the system under consideration we
calculate the phase diagram shown in the plane $(r,\kappa)$ in Fig.\ref{map}.
\begin{figure}[!b]
\centering
\includegraphics[width=70mm]{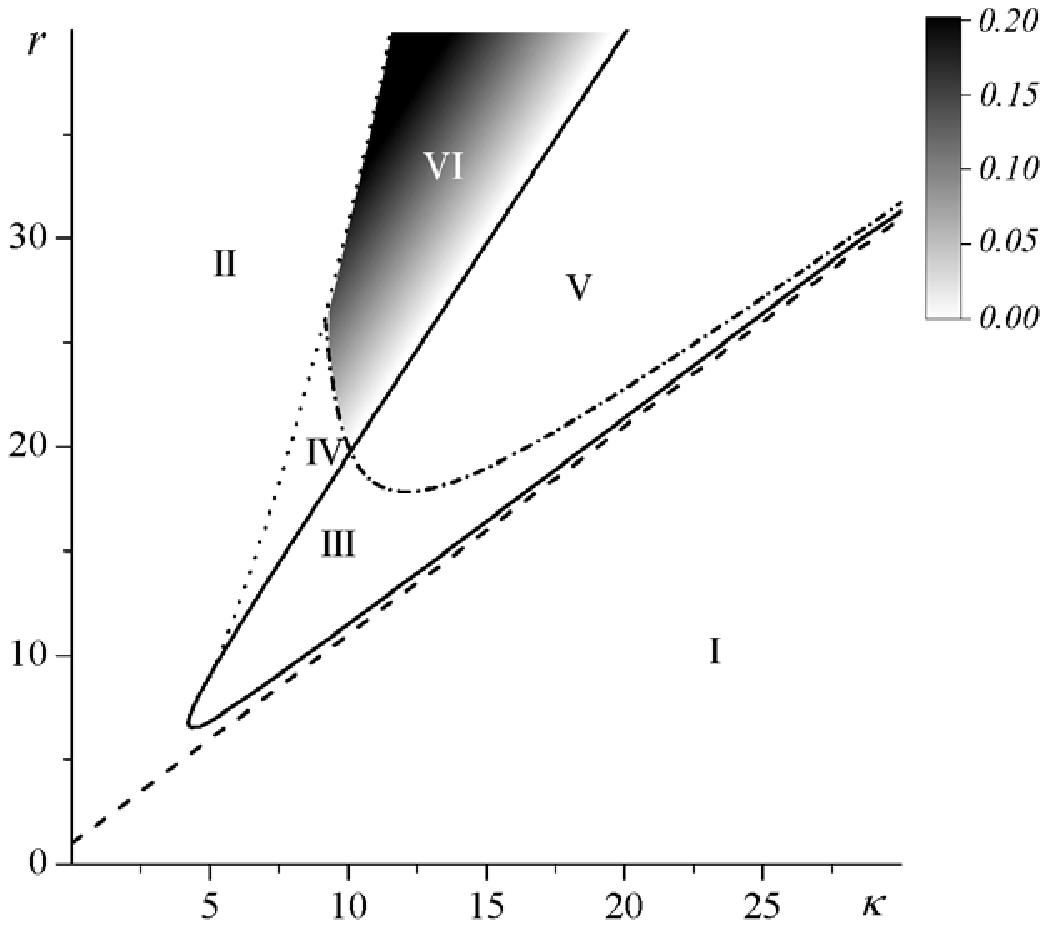}
\caption{Phase diagram of the system (\ref{lor3eq})} \label{map}
\end{figure}
Here the dashed line $r=\kappa+1$ divides the plane $(r,\kappa)$ into two
domains denoted as I and II and relates to the critical values of $r$ and
$\kappa$. Inside the domain I only node is realized. The corresponding phase
trajectories are shown in Fig.\ref{phase}a. Above the dashed line a set of
different dynamical regimes can be observed. Inside the domain II the node
point transforms into a saddle point and two additional stable focuses appear
(see Fig.\ref{phase}b), the corresponding Lyapunov exponents has negative real
parts, and equivalent imaginary ones. In the domain bounded by the solid line
stable limit cycles are formed. This line is obtained as a solution of
Eq.(\ref{eq19}) with assumption that dynamical solutions of the system
Eq.(\ref{lor3eq}) are periodic, i.e. $\Lambda =\pm i\omega$. When we move from
the domain II to the domain III above focuses loss their stability and, as
result, stable limit cycles are realized. All phase trajectories are attracted
by these manifolds. The corresponding limit cycles with unstable focuses inside
of them are shown in Fig.\ref{phase}c. The domain IV bounded by solid, dotted
and dash-dotted lines corresponds to the case where two stable and the
corresponding two unstable limit cycles appear (see Fig.\ref{phase}d). Such a
transformation of dynamical system behaviour (transition from domain III into
domain IV) is a result of double Hopf's bifurcation. Here unstable focuses,
shown in Fig.\ref{phase}c, loss their unstability that leads to appearing
unstable limit cycles inside stable ones (see Fig.\ref{phase}d). When we cross
the dotted line (transition from the domain II into domain IV) the stability of
focuses shown in Fig.\ref{phase}b is not changed, but two stable limit cycles
and two unstable limit cycles appear, as Fig.\ref{phase}d shows. A scenario of
such a bifurcation is as follows. With an increase in one of two system
parameters $r$ or $\kappa$ in a point from the dotted line a new manifold type
of semi-stable limit cycle is formed \cite{Zaslavskiy}. When we move to the
domain IV this semi-stable limit cycle decomposes into the both outer stable
and inner unstable limit cycles. Moving in opposite direction (IV $\to$ II),
one of the well known type of Hoph bifurcation occurs: an annihilation of both
stable and unstable limit cycles.

Let us consider a change of the system behaviour when we move from the domain
III to the domain V bounded by solid and dash-dotted lines. If the dash-dotted
line is crossed, then doubling period bifurcation occurs. Here two stable limit
cycles with unstable focuses, shown in Fig.\ref{phase}c, are transformed into
the one limit cycle (Fig.\ref{phase}e), whose period equals two periods of the
one cycle, shown in Fig.\ref{phase}c. The behaviour of the system inside the
domain V is the same as shown in Fig.\ref{phase}e. If we move from the domain V
to the domain VI we get the period doubling scenario of chaos formation (see
Fig.\ref{phase}f). If the line that divides both domains V and VI is crossed,
then the stability of focuses shown in Fig.\ref{phase}e is changed. It leads to
a formation of two unstable limit cycles and the additional doubling period
bifurcation. If we plunge into the dark zone the phase trajectories become
irregular (see Fig.\ref{phase}g) due to the period doubling bifurcations. In
such a case we obtain the chaotic attractor bounded by outside surface.
Properties of such an attractor will be studied below. If we move from the
domain IV to the domain VI the standard doubling period bifurcation occurs (see
transition from Fig.\ref{phase}d to Fig.\ref{phase}f). The transition from the
domain II into the domain VI is described by the following scenario. In the
line that separates both domains II and VI a complex situation is realized.
Here transverse intersections of both stable and unstable manifolds of the
hyperbolic stationary point (saddle) occur. This indicates a homoclinic
structure and stochastic layer appearance. In such a case the phase space is
characterized by two unstable limit cycles and irregular behavior of
trajectories outside of them (see Fig.\ref{phase}g).

\begin{figure}[!t]
\begin{center}
 \includegraphics[width=75mm]{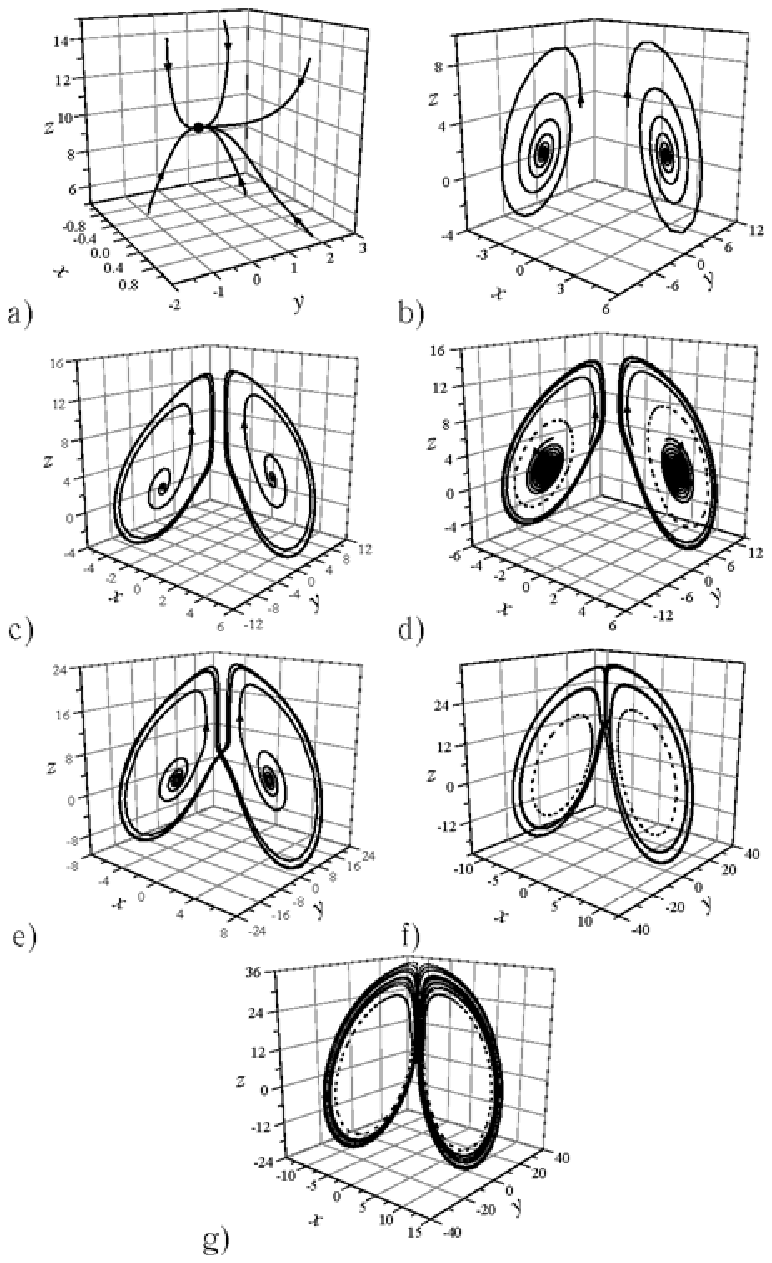}
\caption{Phase portraits of the system (\ref{lor3eq}) at : a) $r=10.0$,
$\kappa=15.0$; b) $r=15.0$, $\kappa=5.0$; c) $r=15.0$, $\kappa=10.0$; d)
$r=20.0$, $\kappa=9.0$; e) $r=25.0$, $\kappa=15.0$; f)~$r=35.0$, $\kappa=12.0$;
g) $r=39.0$, $\kappa=11.5$} \label{phase}
\end{center}
\end{figure}

To elucidate that we deal with irregular behaviour let us use a quantity $h$
known as Kolmogorov-Sinai entropy defined as follows
\begin{equation}
 h=\lim\limits_{
 \begin{scriptsize}
  \begin{array}{l}
   \delta(\tau_0)\to 0\\
   \tau\to\infty
  \end{array}
 \end{scriptsize}
 }\frac{1}{\tau}\ln\left[\frac{\delta(\tau)}{\delta(\tau_0)}\right].
\nonumber
\end{equation}
This quantity allows to observe behaviour of two trajectories 1 and 2
originating at $t=\tau_0$ from two starting points $\vec{u}_1(\tau_0)$ and
$\vec{u}_2(\tau_0)$. It is well known that if $h>0$ then the system manifests
chaotic motion. The corresponding calculations of the separation
$\delta(\tau)=|\vec{u}_2(\tau)-\vec{u}_1(\tau)|$ between these trajectories
versus time $\tau$ when the trajectories are in attractor is shown in
Fig.\ref{distance}.
\begin{figure}[!t]
\centering
\includegraphics[width=70mm]{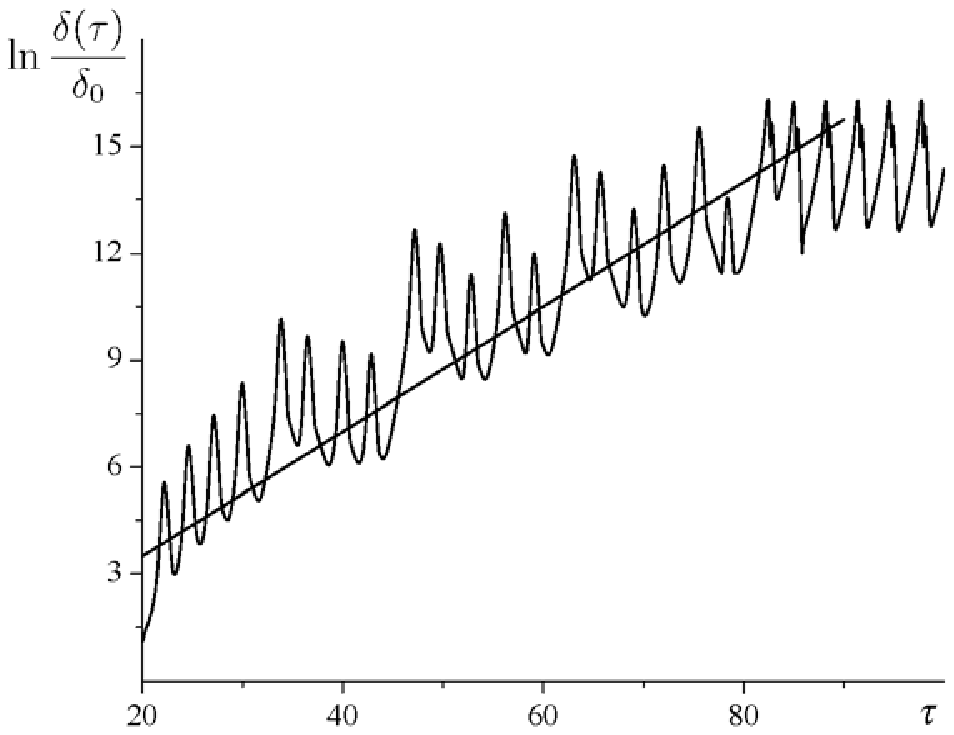}
\caption{The time evolution of a distance $\delta(\tau)$ between two nearest
trajectories in the attractor Fig.\ref{phase}g at
$\delta(\tau_0)=10^{-5}\sqrt{3}$} \label{distance}
\end{figure}
Defining the average inclination angle for the fluctuating curve, one can
define Kolmogorov-Sinai entropy $h=0.1749$. According to the definition of $h$
the characteristic mixing time for the system $t_{mix} \equiv h^{-1} = 5.7175$
in the time units. At $t\ll t_{mix}$ the behaviour of a system is well
predicted, whereas at $t\gg t_{mix}$ only probabilistic description of the
system dynamics is justified. From Fig.\ref{distance} it follows that our
attractor has a bounded size due to that distance $\delta(\tau)$ saturates at
large time $\tau$.

Despite the value of $h$ might also depend on the choice of initial points of
the trajectories it gives not only the qualitative, but also a quantitative
specification of the dynamical regime. The more general criterion which takes
into account the direction of the vector of the initial shift
$\vec\delta(\tau_0)$ is the maximal (global) Lyapunov exponent defined in the
form
\begin{equation}
\Lambda_m\equiv
\Lambda(\vec\delta(\tau_0))=\overline{\lim_{T\to\infty}}\frac{1}{T}\ln \left\|
\frac{\delta {\vec u(\tau)}}{\delta \vec u(\tau_0)}\right\| \nonumber
\end{equation}
where an upper limit and the norm $\|{\vec u}\|$ are used. The quantity
$\Lambda_m$ allows to consider a divergence of any two initially close
trajectories in phase space starting from points $\vec u(\tau)$ and $\vec
u(\tau ')$. It follows, that the divergence of such trajectories is given by
the dependence $\delta\vec{u}(\tau)=\delta\vec{u}(\tau_0)e^{\Lambda_m \tau}$.
It is seen, that at $\Lambda_m<0$ trajectories of dynamical system converge and
hence we get the regular dynamics, where the trajectories moves toward fixed
points in the phase space. In the limit case $\Lambda_m=0$ one gets the limit
cycle in the phase space. If $\Lambda_m>0$ trajectories diverge and the system
shows the chaotic regime. To determine the maximal Lyapunov exponent we use
G.Benettin's algorithm \cite{lichtenberg}. The corresponding spectrum of
maximal Lyapunov exponents $\Lambda_m$ and the domain of its magnitudes are
shown in Fig.\ref{map}.

To prove that above irregular behaviour of phase trajectories can be understood
as chaos let us describe in details a behaviour of the maximal Lyapunov
exponent $\Lambda_m$ versus the pump intensity $r$ at fixed $\kappa$. The
corresponding graph is shown in Fig.\ref{L(se)}.
\begin{figure}[!t]
\centering
\includegraphics[width=70mm]{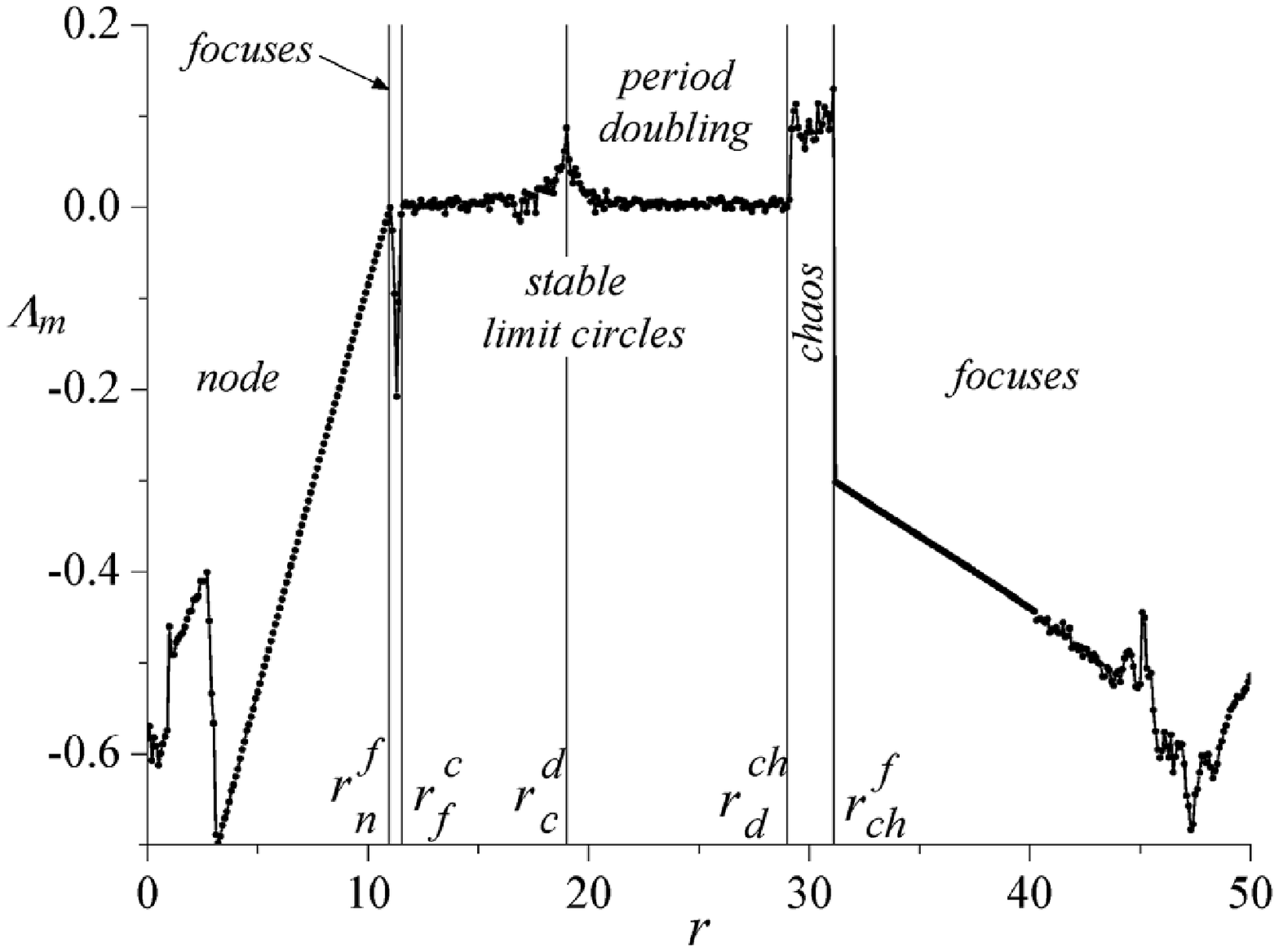}
\caption{Dependence of the maximal Lyapunov exponent vs. pumping $r$ at
$\kappa=10.0$} \label{L(se)}
\end{figure}
It is seen that in the domain of the pump intensity $r\in[0,r^f_n]$ one gets
$\Lambda_m<0$. It means that in the phase space only node point is realized. If
one chooses $r\in(r^f_n,r^c_f]$, then two stable focuses divided by the saddle
point are realized with $\Lambda_m<0$. In the domain $r\in(r^c_f,r^{ch}_d]$
stable focuses transform into stable limit circles. One needs to note that at
$r^d_{c}<r\leq r^{ch}_{d}$ the doubling period bifurcation takes place. At the
such bifurcation point the theoretical value for the maximal Lyapunov exponent
is $\Lambda_m=0$. However, in computer simulations $\Lambda_m\ne 0$, here a
strong nonlinearity in the vicinity of the saddle point acts crucially and
hence the obtained values $\Lambda_m>0$ should be understood as an artifact of
computer simulations \footnote{To prove this we compare a similar doubling
period bifurcation in Roessler system and have found that $\Lambda_m$ abruptly
changes it values in the vicinity of such bifurcation point.}. The critical
value $r^d_{c}$ of the period doubling bifurcation in our system is obtained
when the maximal Lyapunov exponent takes marginal value inside the above
domain. In the interval $r\in(r^{ch}_{d},r^{f}_{ch}]$ we get $\Lambda_m>0$ that
corresponds to divergence of phase trajectories and leads to chaotic regime
formation. At $r>r^{f}_{ch}$ the phase space is characterized by stable
focuses.

\section{Properties of the chaotic regime}

Let us consider statistical properties of irregular behaviour of the system,
which corresponds to the phase portrait in Fig.\ref{phase}g in details. It is
well known that a chaotic regime is characterized by fast decay of an
auto-correlation function and continuous character of frequency spectrum. To
show that our attractor acquires such properties let us calculate an
auto-correlation function $C(\tau)=(2/T)\int_0^{T/2}x(t)x(t+\tau){\rm d}t$ of
the process $x(t)$ shown in Fig.\ref{E(t)}. The auto-correlation function is
presented in Fig.\ref{c(t)}.
\begin{figure}[!t]
\centering
\includegraphics[width=70mm]{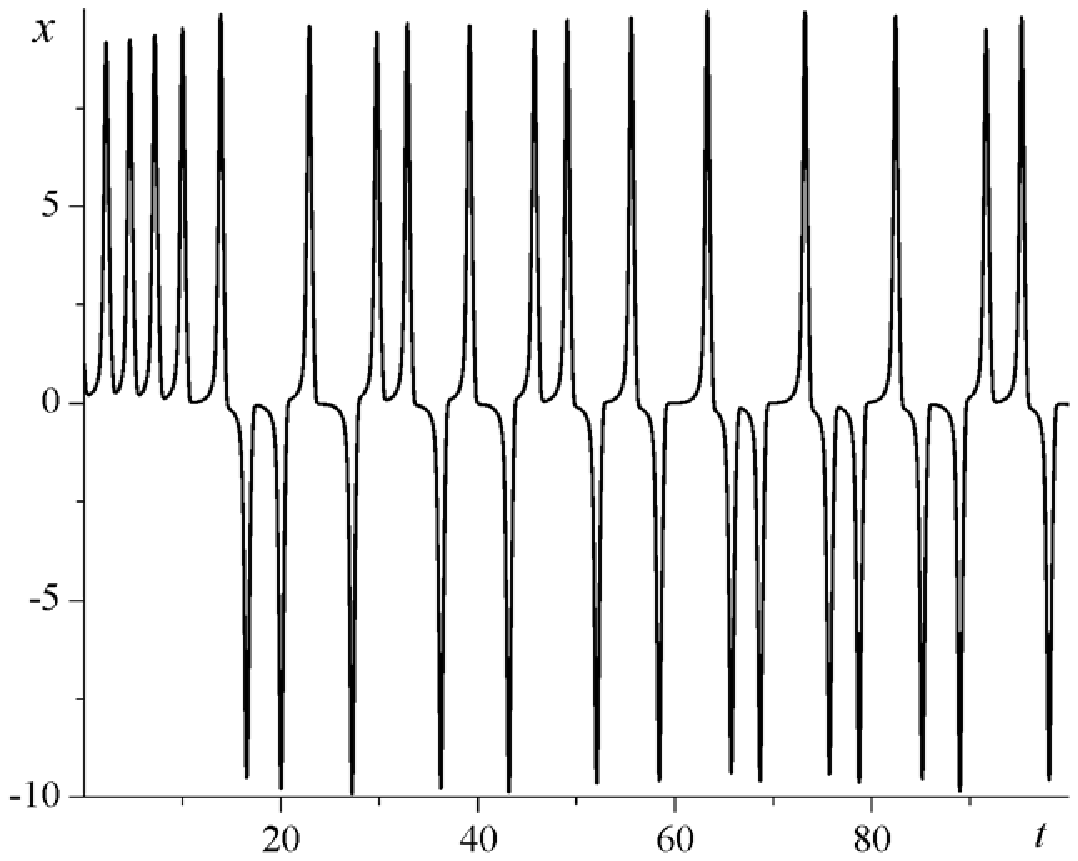}
\caption{Time dependence $x(t)$ at $r=39.0$, $\kappa=11.5$ from
Fig.\ref{phase}g} \label{E(t)}
\end{figure}
\begin{figure}[!t]
\centering
\includegraphics[width=70mm]{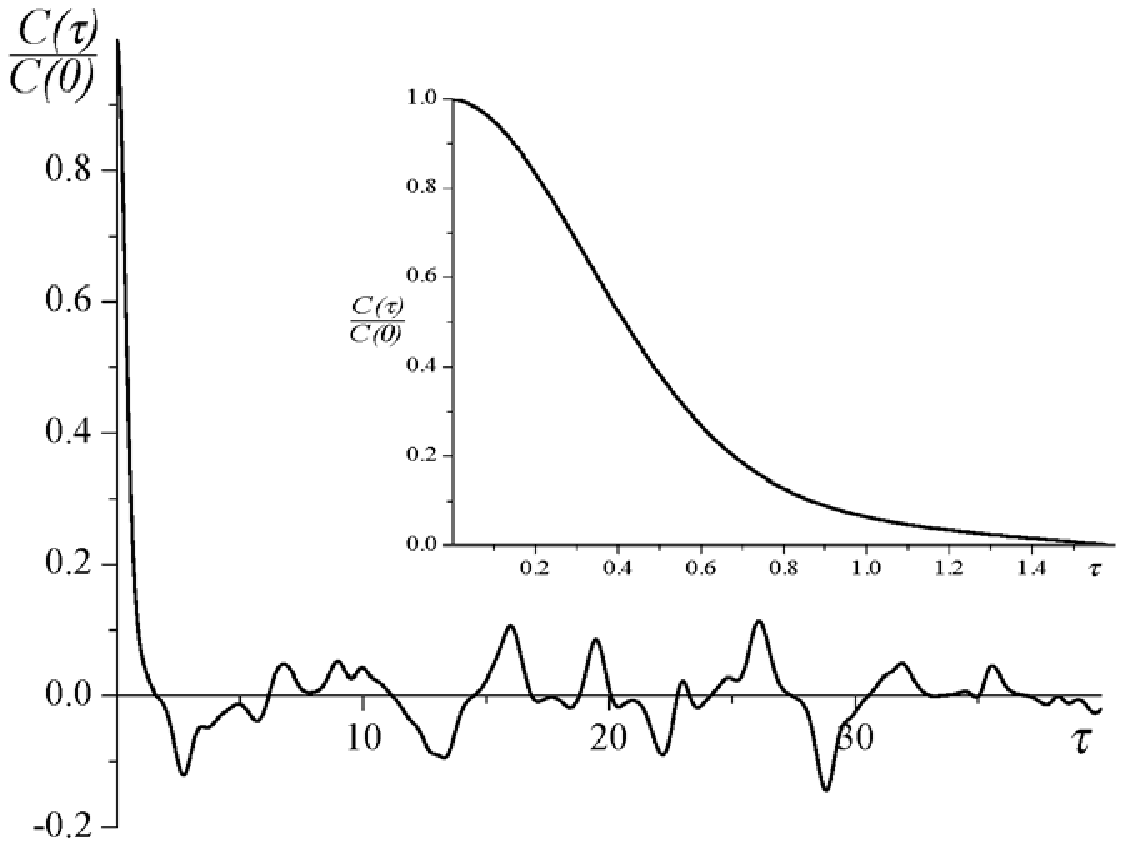}
\caption{Auto-correlation function of process $x(t)$ at $r=39.0$,
$\kappa=11.5$. The corresponding dependence at small time is presented in the
insertion} \label{c(t)}
\end{figure}
It is seen, that auto-correlation function $C(\tau)$ has fast falling down
character as in chaotic systems. It means that the system losses its memory
very quickly. Another criterion to show the chaotic regime is a continuous
frequency spectrum. To obtain it let us use the Fourier transformation of the
correlation function that gives power spectrum function
$S(\omega)=\sqrt{2/\pi}\int_0^T C(\tau)e^{i\omega\tau}{\rm d}\tau.$ The
corresponding frequency spectrum is shown in Fig.\ref{as(w)}.
\begin{figure}[!t]
\centering
\includegraphics[width=70mm]{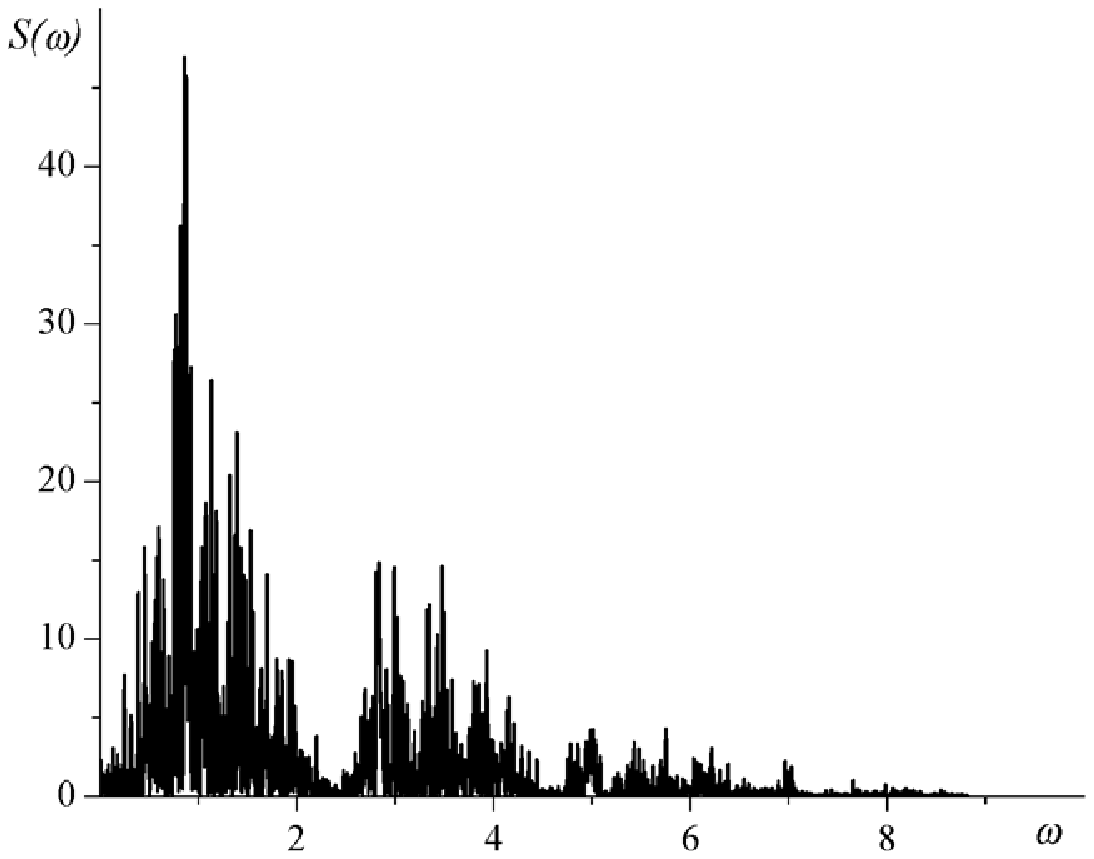}
\caption{Spectral density $S(\omega)$ for time series in Fig.\ref{E(t)}}
\label{as(w)}
\end{figure}
It is seen, that spectral density function has a continuous character. Such a
picture is typical for chaotic regimes \cite{shuster}.

Finally, considering chaotic systems it is interesting to find a
one-dimensional law that allows to represent genuine chaotic behaviour. To this
end we calculate the first return map $z_{n+1}=f(z_n)$ by fixing the
consecutive values of the maxima of the function $z(t)$. Mathematically these
maximal values are $z$-coordinates of points in the Poincare cross-section of
the surface $(r-z)-xy=0$ ($\dot{z}=0$ due to the third equation in the
Eq.(\ref{lor3eq})). The corresponding one-dimensional law $z_{n+1}=f(z_n)$ is
shown in Fig.\ref{sots}.
\begin{figure}[!t]
\centering
\includegraphics[width=70mm]{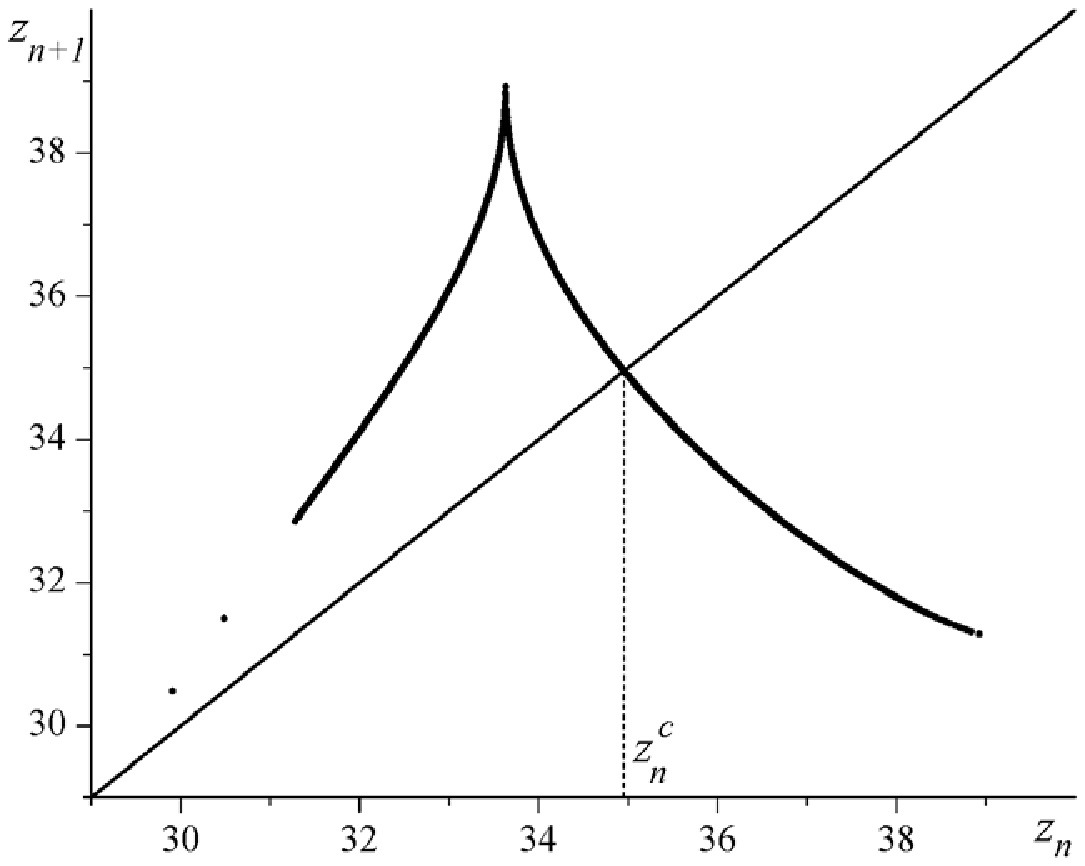}
\caption{The one-dimensional law $z_{n+1}=f(z_n)$ for the system (\ref{lor3eq})
at $r=39.0$, $\kappa=11.5$} \label{sots}
\end{figure}
\begin{figure}[!t]
\centering
 a)\includegraphics[width=70mm]{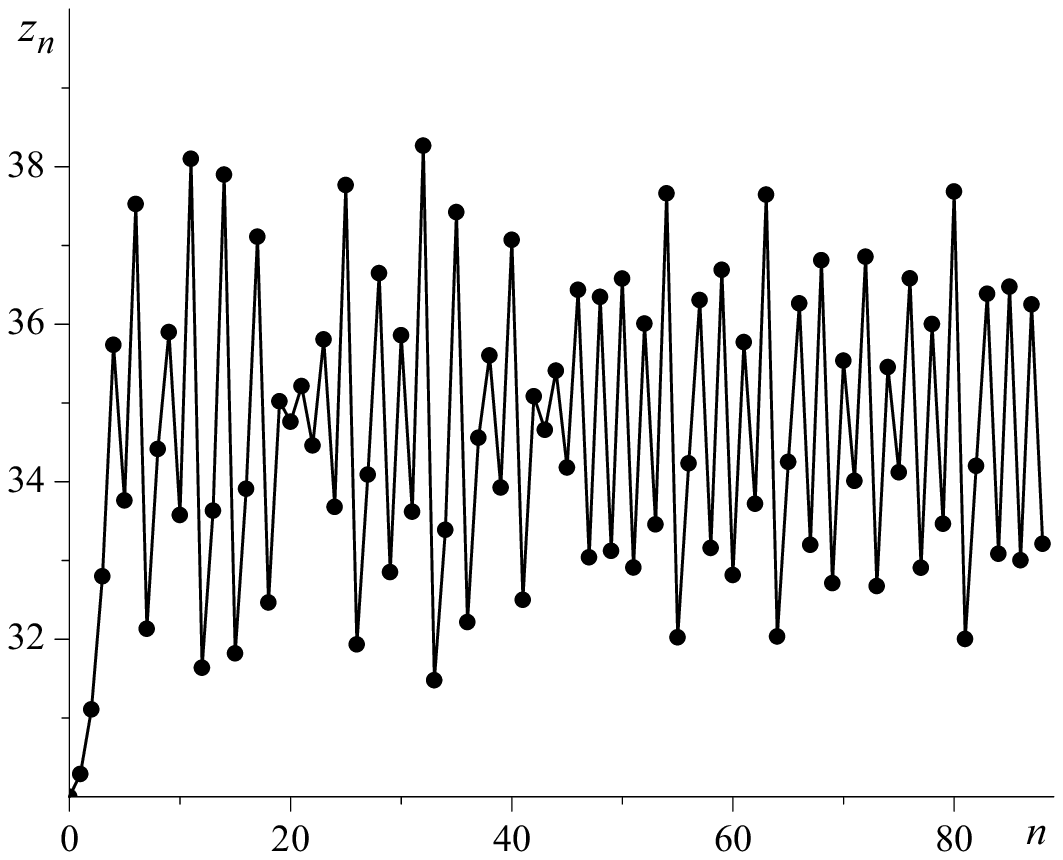}\\
 b)\includegraphics[width=70mm]{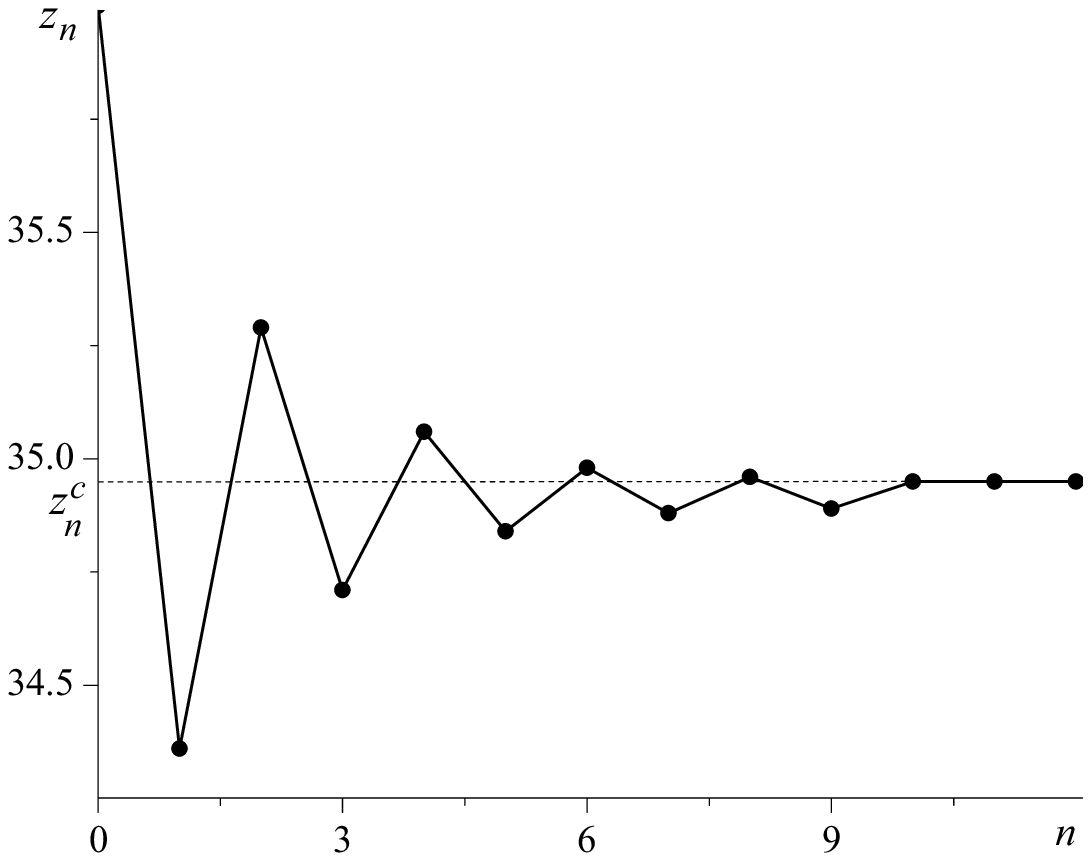}
\caption{Dependence $z_n(n)$ for the different initial values $z_1$ with
$z_n^c=34.95$ at $r=39.0$, $\kappa=11.5$: (a) $z_1<z_n^c$, $z_1=30.0$; (b)
$z_1>z_n^c$, $z_1=36.0$} \label{sotn}
\end{figure}

It is principally that points $(z_n,z_{n+1})$ lie on the one-dimensional curve
with an acute peak with a good accuracy. From the figure Fig.\ref{sots} it
follows that if the initial value $z_1<z_n^c$, then the chaotic character of
dependence $z_n(n)$ is observed. The corresponding dependence $z_n$ versus
number of returns is shown in Fig.\ref{sotn}a, where chaotic character is
visually seen. In the opposite case, when $z_1\geq z_n^c$ trajectories will
converge to the fixed point $z_n^c$ as Fig.\ref{sotn}b shows.

Other important physical quantity characterizing classical dynamic
stochasticity is the fractal dimension of the attractor. For defining of
fractal dimension we will use algorithm of Grassberger-Procaccia
\cite{prokacha}. Let we have set of state vectors $\vec u_i,\ i=1,2\ldots,N$
corresponding to the successive steps of numeric integration. In our case $\vec
u_i$ is the complete set of variables (\ref{lor3eq}) with the values
corresponding to the moments of time $t=t_i$. Then we can use numeric data for
estimation of the following expression
\begin{equation}
C(\ve)=\lim\limits_{N\to\infty}\frac{1}{N(N-1)}\sum\limits_{i,j=1}^N\theta(\ve-\|\vec
u_i-\vec u_j\|)\nonumber
\end{equation}
where $\theta(\varphi)$ is a step function. According to the
Grassberger-Procaccia method, fractal (correlation) dimension of the attractor
may be defined as
$$D_c=\lim\limits_{\ve\to 0}\frac{\log\ C(\ve)}{\log(\ve)}.$$ %
An expected dependence of $C(\ve)$ is $\ve^{D_c}$. So, the corresponding plot
in double logarithmic scale must be a line with angular coefficient $D_c$.
Results of numeric calculations are presented in Fig.\ref{CotVe}.
\begin{figure}[!t]
\centering
 \includegraphics[width=70mm]{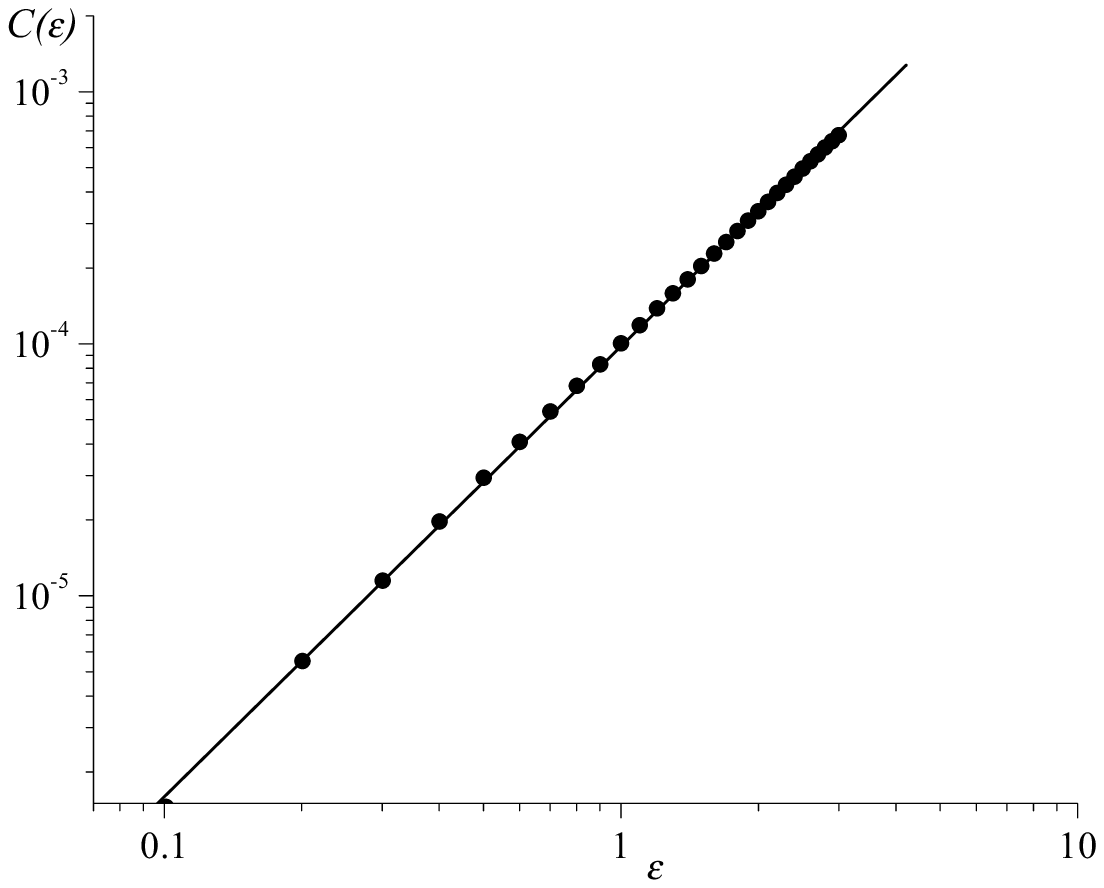}
\caption{Dependence of $C(\ve)$ on the values of $\ve$, plotted by numerical
integration of the set of equations (\ref{lor3eq}) at $r=39.0$, $\kappa=11.5$}
\label{CotVe}
\end{figure}
A solid line corresponds to least-squares approximation of the results of data
processing with $D_c\approx1.84\pm 0.01$. Following Ruelle and Takens
\cite{cite421} one can say that our attractor can be identified as a strange
chaotic attractor.

\section{Conclusions}

We have studied different types of dynamical regimes for the three-component
system with influence of pumping and nonlinear dissipation. It was shown, that
system under consideration is related to absorptive optical bistability
systems. Using Lyapunov exponents approach we have shown, that varying in
system parameters different types of of stable dynamic behaviour of the system
are realized. It was found, that controlling the pumping and/or properties of
additional nonlinear force, related to absorptive medium in optical systems,
unstable and different types of stable dissipative structures can be formed.
Considering dynamics in the three-dimensional phase space a set of bifurcations
from the limit cycle is studied in details.

It was shown, that in the system under consideration a strange chaotic
attractor is realized only if additional nonlinear medium is introduced into
the optical cavity. Properties of such a chaotic attractor were investigated
using Kolmogorov-Sinai entropy, maximal Lyapunov exponents map, spectrum of the
attractor, one-dimensional first return map, and fractal analysis. It was
shown, that two neighbour trajectories in the attractor are divergent with
positive Kolmogorov-Sinai entropy. The corresponding frequency spectrum is
continuous, the temporal auto-correlation function decays fast to zero. From
the first return map it follows, that there is a unique initial point exist
that separates regular and chaotic behaviour of the system. At last, a fractal
(correlation) dimension of the attractor is $D_c\approx1.84\pm 0.01$. From
above criteria on can conclude that obtained attractive manifold can be
considered as a strange attractor as follows from Ruelle and Takens work
\cite{cite421}. Obtained chaotic regime can be controlled by nonlinear
dissipation rate and pumping. Varying such parameters we can obtain a stable
focus and stable limit cycle from the chaotic regime.

Our results are in good correspondence with well known results in systems with
absorptive optical bistability and related to real experimental data for lasers
of ``C''-class \cite{Hanin,kachmarek}. It can be applied to study chaotic
regimes in different kinds of dynamical system that can represent
self-organization processes in models self-consistently described by
hydrodynamic amplitude, conjugated field and control parameter in the form
presented by the Lorenz-Haken model.

\end{document}